\documentclass[12pt]{article}
\pdfoutput=1
\usepackage[utf8]{inputenc}
\usepackage{amsmath}
\usepackage{amssymb}
\usepackage{units}
\usepackage{graphicx}
\usepackage{slashed}
\usepackage{array}
\usepackage[leftcaption,raggedright]{sidecap}
\usepackage{caption}
\usepackage[hidelinks]{hyperref}

\newcommand\pubnumber{NuPhys2023-In\'acio-Parker-Tam}
\newcommand\pubdate{\today}

\def\oxford{University of Oxford}

\def\rs{\footnote{
  Royal Society
}}
\def\stfc{\footnote{
  Science and Technologies Facilities Council
}}
\def\nserc{\footnote{
  Natural Sciences and Engineering Research Council of Canada
}}

\def\Title#1{\begin{center} {\Large #1 } \end{center}}
\def\Author#1{\begin{center}{ \sc #1} \end{center}}
\def\Address#1{\begin{center}{ \it #1} \end{center}}

\newcommand\pubblock{\rightline{\begin{tabular}{l} \pubnumber\\
         \pubdate  \end{tabular}}}
\newenvironment{Abstract}{\begin{quotation}  }{\end{quotation}}
\newenvironment{Presented}{\begin{quotation} \begin{center} 
             PRESENTED AT\end{center}\bigskip 
      \begin{center}\begin{large}}{\end{large}\end{center} \end{quotation}}

\def\beq{\begin{equation}}
\def\eeq#1{\label{#1}\end{equation}}
\def\eeqn{\end{equation}}

\def\beqa{\begin{eqnarray}}
\def\eeqa#1{\label{#1}\end{eqnarray}}
\def\eeqan{\end{eqnarray}}

\let\bar=\overbar

\def\Dslash{\not{\hbox{\kern-4pt $D$}}}
\def\dslash{\not{\hbox{\kern-2pt $\del$}}}

\def\msb{{\bar{\ssstyle M \kern -1pt S}}}

\begin{document}
\begin{titlepage}
\pubblock

\vfill
\Title{The SNO+ Journey to $0\nu\beta\beta$}
\vfill
\Author{ Ana Sofia In\'acio\rs}
\Author{ William Parker\stfc}
\Author{ Benjamin Tam\nserc}
\Address{\oxford}
\vfill
\begin{Abstract}
SNO+ is a large multipurpose experiment with the ultimate goal of searching for the neutrinoless double beta decay in $^{130}\mathrm{Te}$. After a commissioning phase with water as the target medium, during which acquired data allowed for measurements of solar neutrinos and the detection of reactor antineutrinos, SNO+ is now filled with 780 tonnes of liquid scintillator. The higher light yield of the scintillator enhances the physics capabilities of the experiment, and a physics program including reactor, geo and solar neutrinos is currently underway. The water and unloaded scintillator phases provide crucial commissioning milestones in preparation for the tellurium loading, such as calibrating the detector and making extensive background constraint measurements as components of the final scintillator cocktail are gradually added. In a first phase, 3900~kg of natural tellurium (0.5\%) will be added to the scintillator for a predicted sensitivity of about $2\times20^{26}$~years (90\% CL) with 3 years of livetime. Higher tellurium loading will follow for predicted sensitivities above $1\times10^{27}$ years (3\% loading).
\end{Abstract}
\vfill
\begin{Presented}
NuPhys2023, Prospects in Neutrino Physics\\
King's College, London, UK,\\ December 18--20, 2023
\end{Presented}
\vfill
\end{titlepage}
\def\thefootnote{\fnsymbol{footnote}}
\setcounter{footnote}{0}

\section{The SNO+ Experiment}

The SNO+ experiment is a low background kilotonne-scale neutrino experiment located at the SNOLAB facility in Canada. The successor to the Sudbury Neutrino Observatory (SNO) \cite{sno}, SNO+ has inherited much of the infrastructure, with most aspects upgraded and refurbished to accommodate an expanded physics programme utilising liquid scintillator \cite{detector}.

The main body of the SNO+ detector is a 6-m radius acrylic vessel (AV) holding the main target medium. Events within the AV are observed with 9362 inward-facing 8" photomultiplier tubes (PMTs). Mounted on an 8.9\,m radius steel PMT support structure (PSUP) that encompasses the AV, each PMT is attached with a 27-cm diameter concentrator providing a 54\% effective photocoverage.
To mitigate external backgrounds, the detector has a N$_2$ cover gas system, 7000\,m$^3$ of water shielding, and is situated under a 2070\,m rock overburden (corresponding to 6010\,m.w.e.) leading to a muon rate of 0.286$\pm$0.009\,$\mu$/m$^2$/d \cite{muon}. A complete description of the SNO+ detector and all hardware aspects of the experiment is published in \cite{detector}.



SNO+ plans to operate in three phases distinguished by the target medium deployed in the AV.  During an initial ``water phase'', the detector was filled with 905\,tonnes of ultrapure water and operated as a water Cherenkov detector from May 2017 - July 2019. The water was then replaced with 780\,tonnes of liquid scintillator consisting of linear alkylbenzene (LAB) doped with 2.2~g/L of 2,5-diphenyloxazole (PPO) which acts as a fluor. This change in medium increased the light yield of the detector by a factor of $\sim$50, thereby allowing for the study of lower energy processes at higher resolution \cite{scint}. The ``scintillator phase'' is ongoing, having started in April 2022. Due to a break in scintillator fill operations due to the COVID-19 pandemic, a ``partial fill phase," where the detector was filled with 365 tonnes of LAB and 0.6~g/L PPO, had also occurred between March and October 2020.

In the upcoming ``tellurium phase'', the scintillator will be loaded with 3900~kg of $^{\mathrm{nat}}$Te, enabling a search for neutrinoless double beta decay ($0\nu\beta\beta$). The tellurium phase scintillator will also be doped with butylated hydroxytoluene (BHT), 1,4-bis(2-methylstyryl)benzene (bis-MSB), and n,n-dimethyldodecylamine (DDA) which act as an antioxidant, wavelength shifter, and stabilising agent, respectively.

The physics goals of the water phase included a search for invisible nucleon decay, solar neutrinos, and reactor antineutrinos. In the scintillator phase, these goals are expanded to also include geoneutrinos, supernova neutrinos, and a search for dark matter candidates and axion-like particles. The tellurium phase will feature a search for $0\nu\beta\beta$ in addition to continuing the scintillator phase physics programme.

\section{The $0\nu\beta\beta$ Programme}

The search for $0\nu\beta\beta$ using $^{130}\mathrm{Te}$ is the primary objective of the SNO+ experiment. With a $\mathcal{Q}$-value of 2.54\,MeV, the $^{130}\mathrm{Te}$ target was chosen due to the relatively high natural abundance of 34.1\%, thereby precluding enrichment prior to deployment within the detector. This confers the advantage of allowing for a high isotope mass at a low cost compared to other popular $0\nu\beta\beta$ isotopes such as $^{76}\mathrm{Ge}$ or $^{136}\mathrm{Xe}$.

A novel loading technique was developed to deploy the $^{130}\mathrm{Te}$ within the SNO+ liquid scintillator \cite{te}. It was determined that by reacting aqueous telluric acid with 1,4-butanediol, the resultant tellurium butanediol compound would be soluble in the SNO+ liquid scintillator. Furthermore, the isotope mass --- and therefore the $0\nu\beta\beta$ sensitivity of the experiment --- could be increased at relatively low cost.

Utilising this method, an initial amount of 3900~kg $^\mathrm{nat}\mathrm{Te}$ (corresponding to 1300~kg $^{130}\mathrm{Te}$) is planned to be added to the detector in 2025. Increasing the amount of isotope to $\sim12000$~kg $^{nat}\mathrm{Te}$ is planned by 2028, and further scaling of up to $\sim24000$~kg $^{nat}\mathrm{Te}$ has been determined to be possible.

\section{The Background Model}

\begin{figure}
    \centering
    \includegraphics[width=.87\linewidth]{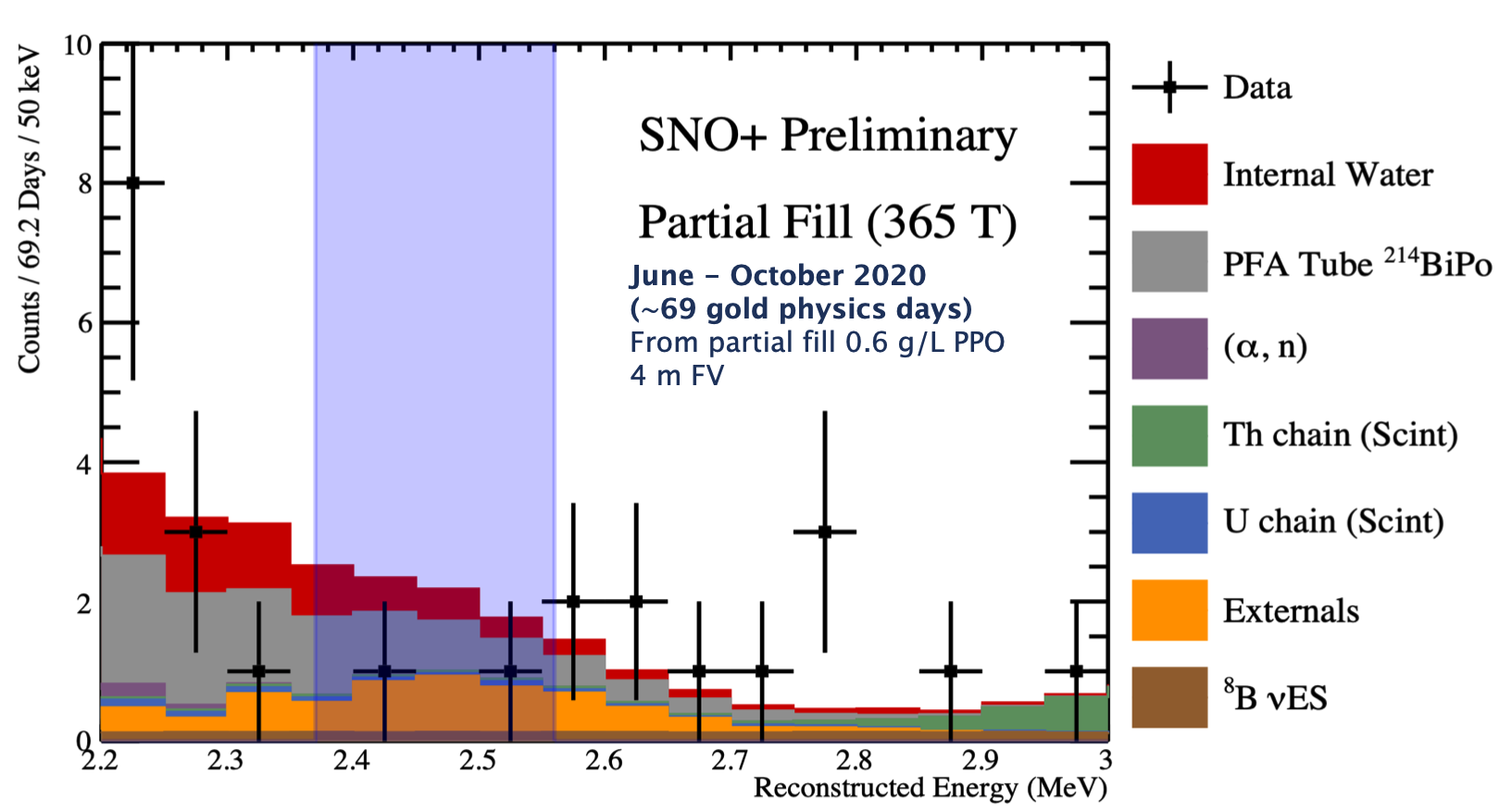}
    \caption{The target out analysis of the partial scintillator phase. The highlighted region is the $0\nu\beta\beta$ ROI. The PFA Tube (grey) was a temporary scintillator fill line that is not present in the scintillator or tellurium phases.}

    \label{fig:targetout}
\end{figure}

Another advantage conferred to the phased approach taken by the SNO+ experiment is the capability to fully understand and quantify the backgrounds in the $0\nu\beta\beta$ energy region of interest (ROI) prior to the deployment of tellurium. This ``target out" analysis allows for the validity of the background model to be directly tested, unforseen scintillator backgrounds to be identified independent of the tellurium systems, and a baseline background level for the ROI to be established. As seen in Figure \ref{fig:targetout}, a target out analysis of the partial fill phase was completed; the target out analysis of the scintillator phase is on-going.

Within the SNO+ experiment, it is expected for there to be six major backgrounds in the ROI, defined as [-0.5, 1.5]$\sigma$ (corresponding to 2.42 -- 2.56 MeV) within a fiducial volume of 3.3~m.
\begin{itemize}
\item The $2\nu\beta\beta$ decay is an inherent standard model process introduced by the isotope and thus an intrinsic background in the $0\nu\beta\beta$ search. This background is expected to contribute 1.21 counts/year in the ROI and can be mitigated through energy resolution.
\item Muon spallation on Te nuclei can result in numerous long lived radioactive backgrounds, including $^{110}\mathrm{Ag}$, $^{60}\mathrm{Co}$, $^{22}\mathrm{Na}$, $^{44}\mathrm{Sc}$, $^{88}\mathrm{Y}$, $^{130}\mathrm{I}$, and $^{130}\mathrm{I}^\mathrm{m}$. These ``cosmogenic" backgrounds have been mitigated by placing the tellurium under the 2070\,m rock overburden of the SNOLAB facility for over 6 years, and can be further mitigated through purification and analytical techniques such as multi-site discrimination. It is expected that the cumulative ROI counts after purification will cause these cosmogenics to contribute 0.11 counts/year to the ROI.
\item $\alpha$-particle captures on the $^{13}\mathrm{C}$ in LAB cause $(\alpha,n)$ interactions. These backgrounds can be mitigated with delayed coincidence tagging, and are expected to contribute 0.02 counts/year to the ROI.
\item The most significant background in the ROI are electrons scattered by $^{8}\mathrm{B}$ solar neutrinos. The flux rate has been well measured by many experiments, including SNO+, and is expected to add 4.61 counts/year to the ROI.
\item $^{238}\mathrm{U}$ and $^{232}\mathrm{Th}$ in the materials that make up the detector can release $\gamma$ particles. These ``external" backgrounds are mitigated by fiducialisation, and were measured in the SNO+ water phase to be 50\% smaller than the target; they are expected to contribute 1.21 counts/year to the ROI.
\item $^{238}\mathrm{U}$ and $^{232}\mathrm{Th}$ within the liquid scintillator and tellurium will also contribute ``internal" background events. Extensive lengths were undertaken to purify the liquid scintillator of these impurities, and the contamination rate of $^{238}\mathrm{U}$ and $^{232}\mathrm{Th}$ have been measured in the liquid scintillator to be $(5.3\pm0.3)\times10^{-17}$\,g/g and $(5.7\pm0.3)\times10^{-17}$\,g/g, respectively. Additional U and Th are expected to be introduced with the deployment of tellurium into the detector, and will be mitigated through further purification. It is expected that these internal backgrounds will contribute a total of 2.22 counts/year to the ROI.
\end{itemize}

\section{Current Status and Outlook}

The SNO+ experiment has begun transitioning from the scintillator to tellurium phases. The deployment of BHT has been completed, and the deployment of bis-MSB is underway. Tracking the $^{210}\mathrm{Po}$ peak have confirmed that the light yield of the scintillator increased by over 50\% following the addition of 0.6\,mg/L bis-MSB; a final bis-MSB concentration of at least 2.2\,mg/L will be added. More reagents will be added to further improve the light yield and stability of the Te-loaded liquid scintillator. 

The ``target out" analysis is on-going to fully quantify the backgrounds prior to the addition of Te. Based on the current understanding of SNO+ backgrounds, the initial projected $0\nu\beta\beta$ sensitivity of the experiment is $S^{0\nu}_{1/2}=2\times10^{26}$\,years after a 3 year live time (90\%C.L.) in an optimised fiducial volume and energy ROI. However, the final sensitivity will depend on the purity achieved during tellurium loading; all Te deployment and purification systems are now in late stages of commissioning. 

\begin{figure}[h]
    \centering
    \includegraphics[width=0.75\linewidth]{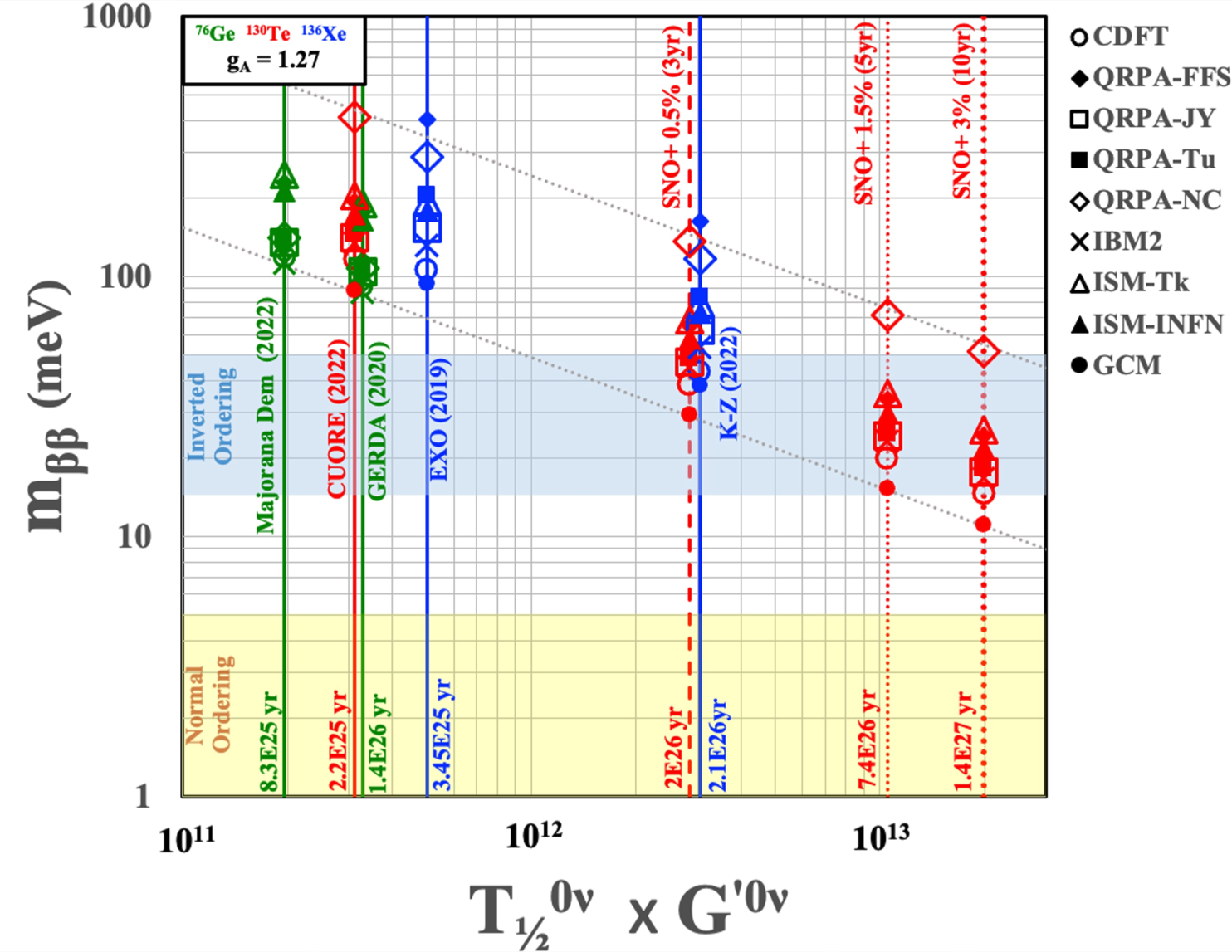}
    \caption{The current bounds (solid), projected near-term sensitivities (dashed) and future sensitivities (dotted) of $0\nu\beta\beta$ experiments.}
    \label{fig:biller}
\end{figure}

The SNO+ experiment aims to deploy an initial amount of 3900~kg $^\mathrm{nat}\mathrm{Te}$ (0.5\% by mass) into the detector in 2025. Furthermore, studies have explicitly determined that the exceptional optical properties and long-term stability of the Te-loaded liquid scintillator applies at concentrations of 3\% by mass. Further loading to this concentration is possible and planned. As can be seen in Figure \ref{fig:biller}, this will allow the SNO+ experiment the potential to probe the entire inverted ordering band.

\pagebreak
\section*{Acknowledgements}
This work is supported by ASRIP, CIFAR, CFI, DF, DOE, ERC, FCT, FedNor, NSERC, NSF, Ontario MRI, Queen’s University, STFC, and UC Berkeley, and have benefited from services provided by EGI, GridPP and Compute Canada. Ana Sofia In\'acio is supported by the Royal Society, William Parker is supported by STFC (Science and Technologies Facilities Council), and Benjamin Tam is supported by NSERC (Natural Sciences and Engineering Research Council of Canada). We thank Vale and SNOLAB for their valuable support.

\providecommand{\href}[2]{#2}\begingroup\raggedright\endgroup

\end{document}